\documentclass[11pt]{article}
\usepackage{a4wide}

\usepackage[dvipdfmx,hiresbb]{graphicx} 
\usepackage{mediabb}                    

\usepackage{amsmath,amssymb}
\usepackage{color}
\usepackage[bookmarks,bookmarksnumbered]{hyperref}
\usepackage{cellspace}
\usepackage{tikz}
\usetikzlibrary{fadings}
\usetikzlibrary{patterns}
\usetikzlibrary{shadows.blur}
\usetikzlibrary{shapes}
\newcommand{\aln}[1]{\begin{align}#1\end{align}}

\begin{document}
\title{\vbox{
\baselineskip 14pt
\hfill \hbox{\normalsize KEK-TH-2449
}}  \vskip 1cm
\bf \Large Baby universes in 2d and 4d theories of quantum gravity
\vskip 0.5cm
}
\author{
Yuta Hamada,\thanks{E-mail: \tt yhamada@post.kek.jp}\,
Hikaru~Kawai,\thanks{E-mail: \tt hikarukawai@phys.ntu.edu.tw}\, and
Kiyoharu Kawana,\thanks{E-mail: \tt kiyoharukawana@gmail.com}
\bigskip\\
\normalsize
\it 
 $^*$  Theory Center, IPNS, KEK, 1-1 Oho, Tsukuba, Ibaraki 305-0801, Japan, \\
\it 
 $^*$  Graduate University for Advanced Studies (Sokendai), \\ \it 1-1 Oho, Tsukuba, Ibaraki 305-0801, Japan,\\
\it 
 $^*$  Department of Physics, Harvard University, Cambridge, MA 02138 USA,\\
 \it
 $^\dagger$ 
 Department of Physics and Center for Theoretical Physics,
 \\
 \normalsize
\it 
  National Taiwan University, Taipei 106, Taiwan, R.O.C. 
\\ 
\normalsize 
\it 
 $^\dagger$ 
 Physics Division, National Center for Theoretical Sciences,
 \\
 \normalsize
\it 
  Taipei 106, Taiwan, R.O.C.
\\ 
\normalsize 
\it  
$^{\ddag}$  Center for Gravitational Physics and Quantum Information, 
\\
\normalsize \it Yukawa Institute for Theoretical Physics,
\\
\normalsize \it 
Kyoto University, Kitashirakawa Oiwakecho, Sakyo-ku, Kyoto 606-8502, Japan
\smallskip
}
\date{}

\maketitle  
\vspace*{1cm}  
\begin{abstract} 
The validity of the Coleman mechanism, which automatically tunes the fundamental constants, is examined in two-dimensional and four-dimensional quantum gravity theories.
First, we consider two-dimensional Euclidean quantum gravity on orientable closed manifolds coupled to conformal matter of central charge $c \leq1$.
The proper time Hamiltonian of this system is known to be written as a field theory of noncritical strings, which can also be viewed as a third quantization in two dimensions.
By directly counting the number of random surfaces with various topologies, we find that the contribution of the baby universes is too small to realize the Coleman mechanism.
Next, we consider four-dimensional Lorentzian gravity. Based on the difference between the creation of the mother universe from nothing and the annihilation of the mother universe into nothing, we introduce a non-Hermitian effective Hamiltonian for the multiverse.
We show that Coleman's idea is satisfied in this model and that the cosmological constant is tuned to be nearly zero.
Potential implications for phenomenology are also discussed.
\end{abstract} 

\setcounter{page}{1} 

\newpage  

\tableofcontents   

\newpage  

\section{Introduction}
\label{Sec:intro}

The smallness of the cosmological constant is one of the great mysteries of particle physics.
In the late 1980's, Coleman proposed a solution \cite{Coleman:1988tj,Klebanov:1988eh} based on the effects of Euclidean wormholes~\cite{Giddings:1987cg} (see also Refs.~\cite{Kawai:2013wwa,Hebecker:2018ofv} for reviews).
After summing the wormholes, the low energy theory is described by an ensemble average of various coupling constants, including the cosmological constant.
However, Coleman's original proposal has problems such as \cite{Fischler:1988ia,Polchinski:1988ua,Fischler:1989ka}. 
These problems seem to stem from the pathology of the 4d Euclidean gravity associated with the conformal mode.
To overcome this problem, a Lorentzian formulation of Coleman's mechanism was proposed and studied~\cite{Kawai:2011rj,Kawai:2011qb,Hamada:2014ofa,Hamada:2014xra,Hamada:2015dja}.

On the other hand, significant progress has recently been made toward resolving the information paradox of the black hole ~\cite{Penington:2019npb,Almheiri:2019psf}.
At least in two dimensions, the replica wormhole~\cite{Penington:2019kki,Almheiri:2019qdq} plays an important role in reproducing the unitary page curve of the black hole entropy (see Refs.~\cite{Almheiri:2020cfm,Raju:2020smc} for reviews).

Given the importance of wormhole, it is interesting to revisit the Coleman's mechanism in two dimensions.
Indeed, 2d Euclidean quantum gravity on closed manifolds coupled to a matter field with central charge $c\leq1$ is well-defined.
Its proper time Hamiltonian can be regarded as a kind of field theory of noncritical strings.
Thus, the validity of Coleman's proposal can be clearly discussed.\footnote{The analysis of wormholes in the worldsheet theory of critical strings has been done in \cite{Lyons:1991im}.
See also Ref.~\cite{Betzios:2020nry} for recent study of the Liouville theory coupled to $c=1$ matter using the matrix quantum mechanics.
 }

In this paper, we first show that the sum of topologies in 2d Euclidean gravity does not lead to an automatic tuning of the cosmological constant by explicitly counting the number of random surfaces.\footnote{The fluctuation of the cosmological constant in 2d gravity is also considered in \cite{Ambjorn:2021wdm}.} 
We argue that this is true for a wide range of modifications of the 2d Euclidean gravity based on the matrix model.
Next, we consider 4d Lorentzian gravity and introduce an effective Hamiltonian of the multiverse consisting of the creation and annihilation operators of the mother and baby universes. 
The Hamiltonian is non-Hermitian due to the difference between the creation of the mother universe from nothing and the annihilation of the mother universe into nothing.
In this model, the Coleman mechanism is realized and the effective cosmological constant is tuned to almost zero. 

The paper is organized as follows.
In Section~\ref{Sec:Euclidean_2d}, we first outline the path integral formulation of 2d Euclidean gravity (non-critical strings). In particular, we introduce the Hamiltonian formulation for 2d gravity coupled to $(2,q)$ minimal matter.\footnote{It includes the Jackiw-Teitelboim (JT) gravity as a limit $q \rightarrow \infty$~\cite{Teitelboim:1983ux,Jackiw:1984je}.} We then show that the effect of the microscopic baby universes is too small compared to the macroscopic topology changes to realize the Coleman mechanism.
We then consider modifications of 2d gravity based on the matrix model, and discuss
that the Coleman mechanism works in Lorentzian gravity.
In Section~\ref{Sec:Lorentzian}, we consider Lorentzian gravity. The processes of creation and annihilation of the mother and baby universes are investigated, and a non-Hermitian effective Hamiltonian describing a Lorentzian multiverse is introduced.
We show that Coleman's idea is satisfied in this model.
We also discuss the potential implications for phenomenology.

\section{2d Euclidean quantum gravity with various topologies}\label{Sec:Euclidean_2d}
In this section, we examine the possibility of obtaining ensemble averages for the coupling constants from the sum of topologies in two-dimensional gravity.
The Euclidean 2d gravity coupled to a matter field with central charge $c\leq1$ is well-defined without suffering from the problem of the conformal mode.\footnote{This is related to the fact that the number of degrees of freedom is negative in 2d gravity.}
It can be defined either by continuum theory \cite{Knizhnik:1988ak,David:1988hj,Distler:1988jt} or by dynamical triangulation \cite{David:1984tx,Kazakov:1985ds,Boulatov:1986jd,Ambjorn:1985az,Kazakov:1985ea}. In particular, all topologies can be summed using the matrix model~ \cite{Brezin:1990rb,Douglas:1989ve,Gross:1989vs}. (See also  Ref.~\cite{Kawai:1991qv,Ginsparg:1993is,DiFrancesco:1993cyw,Nakayama:2004vk} for reviews. )

As is well known, 4d Euclidean gravity has difficulties due to instability of the conformal mode. Also, whether a microscopic wormhole is more important than a macroscopic topology change depends on the dimension of spacetime. Nevertheless, the 2d Euclidean wormhole is a good clue to investigate the 4d Lorentzian multiverse, as we will see in the next section.

\subsection{Formulations in Continuum theory}
In this subsection, we introduce two formalisms of 2d Euclidean gravity.

\subsubsection{Hamiltonian Formalism: Non-critical string field theory}\label{Sec:non-critical_string}
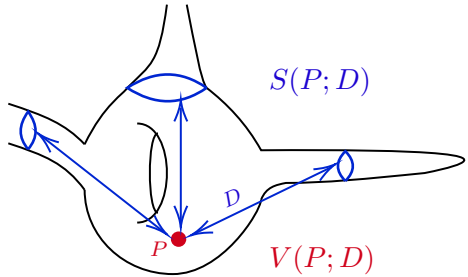
\begin{figure}
    \centering

\tikzset{every picture/.style={line width=0.75pt}} 

\begin{tikzpicture}[x=0.75pt,y=0.75pt,yscale=-1,xscale=1]

\draw  
(267.5,58.5) .. controls (267,64.97) and (247.58,96) .. (223,96) .. controls (198.43,96) and (178.5,76.97) .. (178.5,53.5)  ;
\draw 
(178.5,53.5) .. controls (168.5,40) and (158.33,36) .. (140,30) ;
\draw 
(178.5,30.5) .. controls (168.5,20) and (158.33,16) .. (140,10) ;
\draw 
(178.5,30.5) .. controls (188.5,10) and (198.33,5) .. (200,2) .. controls (212,-10) and (216,-10) .. (220,-40) ;
\draw 
(267.5,58.5) .. controls (270,52) and (275,51) .. (280,50) .. controls (300,50) and (320,48) .. (350,45) .. controls (400,35) and (350,33) .. (267.5,33.5) .. controls (260,20) and (250.33,10) .. (240,2) .. controls (237,-2) and (234,-20) .. (230,-40) ;
\draw 
[shift={(-10,0)}] (215,20) .. controls (235,20) and (235,70) .. (215,70) ;
\draw 
[shift={(-10,0)}] (225.5,25) .. controls (220,25) and (220,65) .. (225.5,65) ;
\draw [color={rgb, 255:red, 0; green, 38; blue, 209 }  ,draw opacity=1 ][line width=1]    (310,49) .. controls (315,44) and (315,39) .. (310,34) .. controls (305,39) and (305, 44) .. (310,49);
\draw [color={rgb, 255:red, 0; green, 38; blue, 209 }  ,draw opacity=1 ][line width=1]    (240,2) .. controls (227,-7) and (214,-7) .. (200,2) .. controls (214,11) and (227, 11) .. (240,2);
\draw [shift={(0,-20)}] [color={rgb, 255:red, 0; green, 38; blue, 209 }  ,draw opacity=1 ][line width=1]    (150,53) .. controls (155,47) and (155,41) .. (150,34) .. controls (145,41) and (145,47) .. (150,53);

\draw  [color={rgb, 255:red, 208; green, 2; blue, 27 }  ,draw opacity=1 ][fill={rgb, 255:red, 208; green, 2; blue, 27 }  ,fill opacity=1 ] (222.5, 78.38) .. controls (222.5, 80.24) and (224.01, 81.75) .. (225.88, 81.75) .. controls (227.74, 81.75) and (229.25, 80.24) .. (229.25, 78.38) .. controls (229.25, 76.51) and (227.74, 75) .. (225.88, 75) .. controls (224.01, 75) and (222.5, 76.51) .. (222.5, 78.38) -- cycle ;

 \draw (270,80) node [anchor=north west][inner sep=0.75pt]    {$\textcolor[rgb]{0.82,0.01,0.11}{V(P;D)}$};
 \draw (210,78) node [anchor=north west][inner sep=0.75pt]    {\textcolor[rgb]{0.82,0.01,0.11}{{\fontsize{8pt}{8pt}\selectfont $P$}}};
\draw (270,-10) node [anchor=north west][inner sep=0.75pt]  {$\textcolor[rgb]{0.11,0.01,0.82}{S(P;D)}$};
\draw (245,55) node [anchor=north west][inner sep=0.75pt] [rotate=24] {\textcolor[rgb]{0.11,0.01,0.82}{{\fontsize{8pt}{8pt}\selectfont $D$}}};

\draw [color={rgb, 255:red, 0; green, 38; blue, 209 }  ,draw opacity=1 ]   (233,76) -- (303,42) ;
\draw [shift={(305,40.5)}, rotate = 150] [color={rgb, 255:red, 0; green, 38; blue, 209 }  ,draw opacity=1 ][line width=0.75]    (10.93,-3.29) .. controls (6.95,-1.4) and (3.31,-0.3) .. (0,0) .. controls (3.31,0.3) and (6.95,1.4) .. (10.93,3.29)   ;
\draw [shift={(232,76.5)}, rotate = -24] [color={rgb, 255:red, 0; green, 38; blue, 209 }  ,draw opacity=1 ][line width=0.75]    (10.93,-3.29) .. controls (6.95,-1.4) and (3.31,-0.3) .. (0,0) .. controls (3.31,0.3) and (6.95,1.4) .. (10.93,3.29)   ;

\draw [color={rgb, 255:red, 0; green, 38; blue, 209 }  ,draw opacity=1 ]   (227,72) -- (227,12) ;
\draw [shift={(227,8)}, rotate = 90] [color={rgb, 255:red, 0; green, 38; blue, 209 }  ,draw opacity=1 ][line width=0.75]    (10.93,-3.29) .. controls (6.95,-1.4) and (3.31,-0.3) .. (0,0) .. controls (3.31,0.3) and (6.95,1.4) .. (10.93,3.29)   ;
\draw [shift={(227,72.5)}, rotate = -90] [color={rgb, 255:red, 0; green, 38; blue, 209 }  ,draw opacity=1 ][line width=0.75]    (10.93,-3.29) .. controls (6.95,-1.4) and (3.31,-0.3) .. (0,0) .. controls (3.31,0.3) and (6.95,1.4) .. (10.93,3.29)   ;

\draw [color={rgb, 255:red, 0; green, 38; blue, 209 }  ,draw opacity=1 ]   (155,25) -- (220,76) ;
\draw [shift={(155,25)}, rotate = 40] [color={rgb, 255:red, 0; green, 38; blue, 209 }  ,draw opacity=1 ][line width=0.75]    (10.93,-3.29) .. controls (6.95,-1.4) and (3.31,-0.3) .. (0,0) .. controls (3.31,0.3) and (6.95,1.4) .. (10.93,3.29)   ;
\draw [shift={(220,76)}, rotate = 220] [color={rgb, 255:red, 0; green, 38; blue, 209 }  ,draw opacity=1 ][line width=0.75]    (10.93,-3.29) .. controls (6.95,-1.4) and (3.31,-0.3) .. (0,0) .. controls (3.31,0.3) and (6.95,1.4) .. (10.93,3.29)   ;

\end{tikzpicture}

    \caption{Illustration of $V(P;D)$ and $S(P;D)$. Here $V(P;D)$ is the set of points whose geodesic distance from $P$ is less than or equal to $D$. The boundary of $V(P;D)$ is denoted by $S(P;D)$. In the figure, $S(P;D)$ consists of three loops.}
    \label{Fig:proper_time}
\end{figure}

In Refs.~\cite{Ishibashi:1993nq, Fukuma:1993np}, a Hamiltonian formalism was proposed for 2d Euclidean gravity, in which the geodesic distance is considered as time (See Ref.~\cite{Kawai:1993cj} for the Hamiltonian in dynamical triangulation). 
This theory can be regarded as a string field theory, since the Hamiltonian describes the creation and annihilation of universes of spatial dimension 1. 
It can also be viewed as a two-dimensional third quantization theory~\cite{Giddings:1988wv}.
This formalism is convenient to generalize to Lorentzian spacetime, which we will explore in Section~\ref{Sec:Lorentzian}.

Let us consider 2d spacetime, and take an arbitrary point $P$ (See Fig. \ref{Fig:proper_time} for illustration).
Then, the set of points $V(P;D)$ is defined as
\begin{align}
V(P;D) = \{ Q\in (\text{spacetime})| d(P,Q)\leq D\}
\end{align}
where $d(P,Q)$ is the geodesic distance between $P$ and $Q$.
Let $S(P;D)$ be the boundary of $V(P;D)$.

Next, we introduce operators $\psi^\dagger(\ell)$ and $\psi(\ell)$, which create and annihilate loops (1d spaces) of length $\ell$, respectively.
They satisfy the relation,
\begin{align}
    [\psi(\ell),\psi^\dagger(\ell')]=\delta(\ell-\ell')~,
\end{align}
and the vacuum (the absence of space) is defined as
\begin{align}
\psi(\ell)|0\rangle=\langle0|\psi^\dagger(\ell)=0.
\end{align}
For simplicity, we take the $(2,q)$ minimal model as the matter field.\footnote{The $(2,q)$ minimal model has
$c = -3q +13- \frac{12}{q}$. 
For example, $q=1,3,\infty$ gives $c=-2, 0, -\infty,$ respectively.
}
In that case, there is no need to introduce any extra degrees of freedom other than $l$.\footnote{Non-critical string field theory for the minimal unitary series $(p,p+1)$ ($p=2,3,\cdots$) is given in Ref.~\cite{Ikehara:1994vx}.}
Then, the state with $k$ loops of length $\ell_1,\cdots,\ell_k$ can be written as
\begin{align}
|\ell_1,\cdots,\ell_k\rangle = \psi^\dagger(\ell_1)\cdots\psi^\dagger(\ell_k)|0\rangle~,
\end{align}
and the state of the boundary surface $S(P;D)$ is represented by their superposition:
\begin{align}
|S(P;D)\rangle = \sum_{k=0}^\infty \int^\infty_0d\ell_1 \cdots \int^\infty_0\,d\ell_k c_k(\ell_1,\cdots,\ell_k)|\ell_1,\cdots,\ell_k\rangle~. 
\end{align}

We can define Hamiltonian~\cite{Kawai:1993cj,Ishibashi:1993nq}, which describes the infinitesimal translation of the proper time $D$.
\begin{align}
\frac{d}{dD} |S(P;D)\rangle = - H_\mathrm{Euclid} |S(P;D)\rangle~,
\label{Eq:Euclidean_Schroedinger}
\end{align}
where
$H_\mathrm{Euclid}$ is given by
\begin{align}
H_\mathrm{Euclid} = &\int^\infty_0 d\ell_1 d\ell_2 \,\psi^\dagger(\ell_1) \psi^\dagger(\ell_2) \psi(\ell_1+\ell_2)
+ \int^\infty_0 d\ell_1 d\ell_2 \,\psi^\dagger(\ell_1+\ell_2) \psi(\ell_1) \psi(\ell_2)
\nonumber\\
&+\int^\infty_0 d\ell \,\rho(\ell) \psi(\ell)~.
\label{Eq:Euclidean_Hamiltonian}\end{align}
The source function $\rho(\ell)$ is\footnote{To be precise, the source term is defined through the Laplace transformation.} 
\begin{align}
\rho(\ell)=
\begin{cases}
\lambda\,\delta(\ell) \quad \text{for $(2,1)$ topological gravity ($c=-2$) \cite{Ishibashi:1993nqz}}\\
\lambda\,\delta(\ell) + \delta''(\ell) \quad \text{for $(2,3)$ pure gravity ($c=0$) \cite{Ishibashi:1993nq}}
\end{cases}.
\end{align}
This function is related to the disk amplitude:
\begin{align}
\tilde{\rho}(\zeta)=
\frac{\partial}{\partial\zeta}(\tilde{D}(\zeta))^2,
\end{align}
where $\tilde{D}(\zeta)$ is the Laplace transformation of the disk amplitude $D(l)$.

The function $\rho$ corresponding to the JT gravity \cite{Teitelboim:1983ux,Jackiw:1984je} can be obtained as follows.
Its action is given by
\begin{align}
S_{JT}=-\frac{S_0}{2\pi}\left[
\frac{1}{2}\int_{\mathcal{M}}\sqrt{g}R + \int_{\partial\mathcal{M}} \sqrt{h} K 
\right]
-\left[ \frac{1}{2}\int_{\mathcal{M}}\sqrt{g}\Phi(R+2)+\int_{\partial\mathcal{M}}\sqrt{h}\Phi(K-1)\right],
\end{align}
where $S_0$ is a constant, $K$ is the boundary extrinsic curvature, and $\Phi$ is the dilaton.
We consider the case where $\mathcal{M}$ has the disk topology. From the variation of $\Phi$, we obtain that the metric is the Euclidean $AdS_2$ ($EAdS_2$).
In the Poincare patch ($ds^2=(d\tau^2+dz^2)/z^2$),  the solution of the dilaton is
\begin{align}
\Phi=\frac{2\pi\gamma}{z}~,
\end{align}
where $\gamma$ is a constant. 
Then $\tilde{D}$ is given by \cite{Saad:2019lba,Hirano:2021rzg}
\begin{align}
\tilde{D}_{JT}(\zeta)=
e^{S_0}\frac{\gamma}{2\pi^2}\sinh\left(2\pi\sqrt{2\gamma \zeta}\right)~.
\end{align}
This leads to
\begin{align}
\tilde{\rho}_{JT}(\zeta)=
\frac{\partial}{\partial\zeta}\tilde{D}_{JT}^2=
e^{2S_0}\frac{\gamma^{5/2}}{4\pi^3}\sqrt{\frac{2}{\zeta}}\sinh\left(4\pi\sqrt{2\gamma \zeta}\right)
=e^{2S_0}\frac{\gamma^2}{\pi^2}\sum_{n=1}^\infty \frac{(2\gamma)^n}{(2n-1)!}\zeta^{n-1}~,
\end{align}
from which we obtain
\begin{align}
\rho_{JT}(l)=e^{2S_0}\frac{\gamma^2}{\pi^2}\sum_{n=1}^\infty \frac{(2\gamma)^n}{(2n-1)!}\delta^{(n-1)}(\ell)~.
\end{align}
Here $\delta^{(n)}(l)$ is the $n$-th derivative of the Dirac delta function.
Thus the JT gravity is obtained as the limit $p=2, q\to\infty$ \cite{Saad:2019lba} (see also Refs.~\cite{Mertens:2019tcm,Mertens:2020hbs,Turiaci:2020fjj,Okuyama:2021eju,Gregori:2021tvs}). 

Using Eq.~(\ref{Eq:Euclidean_Schroedinger}), we can evaluate the length distribution of the circles that consist of the boundary $S(P;D)$~
\cite{Kawai:1993cj,Gubser:1993vx}. 
For example, in the case of $c=0$ ($q=3$), the expectation value of the number of loops with length from $L$ to $L+dL$ contained in $S(P;D)$ is given by
\begin{align}
&n(L;D)dL= \frac{3}{7\sqrt{\pi}D^2} \left(x^{-5/2}+\frac{1}{2}x^{-3/2}+\frac{14}{3}x^{1/2}\right) e^{-x}dL~,
&&x:=\frac{L}{D^2}~.
\end{align}
This implies that a large number of small baby universes will be created. 
Nevertheless, we cannot conclude that the Coleman mechanism is realized, as we will discuss in Section~\ref{Sec:absence}.

\subsubsection{Path Integral Formalism}\label{Sec:DDK}

The Liouville action appears in the quantization of 2d gravity coupled to conformal matter~\cite{David:1988hj,Distler:1988jt} (See also Ref.~\cite{Knizhnik:1988ak}). 
In Section~\ref{Sec:absence}, we use this to examine the possibility of a Coleman mechanism in 2d Euclidean gravity.

We start from the partition function,
\begin{align}
    Z=\int \frac{\mathcal{D}g}{\text{vol(Diff)}}e^{-\frac{\mu_0}{2\pi}\int d^2x \sqrt{g}}Z_M[g]~.
\end{align}
Here $Z_M[g]$ is the partition function of a conformal field with central charge $c$ defined on the background metric $g_{\mu\nu}(x)$, vol(Diff) stands for the volume of the space of diffeomorphisms, and $\mu_0$ is the (bare) cosmological constant.
The path measure $\mathcal{D}g$ is induced from the diffeomorphism invariant norm:
\begin{align}
    ||\delta g||^2 = \int d^2x \sqrt{g}g^{\mu\nu}g^{\lambda\rho}\delta g_{\mu\lambda}\delta g_{\nu\rho}~.
\label{Eq:g_norm}\end{align}
In the conformal gauge, the metric is parametrized as
\begin{align}
    g_{\mu\nu}(x)=\hat{g}_{\mu\nu}(\tau,x)\,e^{\phi(x)}~,
\label{Eq:decomposition}\end{align}
where $\phi$ and $\tau$ are the conformal mode and moduli, respectively.
After this decomposition, the partition function becomes
\begin{align}
    \int d\tau\,\mathcal{D}_1\phi\, \Delta_{FP}[\hat{g}e^\phi]Z_M[\hat{g}e^\phi]e^{-\frac{\mu_0}{2\pi}\int \sqrt{\hat{g}}e^\phi d^2x}
    =\int d\tau\,\mathcal{D}_1\phi\, \Delta_{FP}[\hat{g}]Z_M[\hat{g}]e^{\frac{c-26}{48\pi}S_L[\hat{g};\phi]-\frac{\mu_0}{2\pi}\int \sqrt{\hat{g}}e^\phi d^2x}~,
\label{Eq:Z_conformal_gauge}\end{align}
where $\Delta_{FP}[\hat{g}e^\phi]$ is the Faddeev-Popov determinant, $S_L(\hat{g};\phi)$ is the (unrenormalized) Liouville action, 
\begin{align}
    S_L(\hat{g};\phi)=\int d^2x \sqrt{g}\left(\frac{1}{2}g^{\mu\nu}\partial_\mu\phi\partial_\nu\phi+ R\phi\right)~.
\end{align}
$\mathcal{D}_1\phi$ is the measure induced from the norm
\begin{align}
    ||\delta\phi||^2_1=\int d^2x \sqrt{\hat{g}}\,e^\phi(\delta\phi)^2~,
\label{Eq:phi_norm_1}\end{align}
which is derived from Eq.~\eqref{Eq:g_norm}.
In the last equality of Eq.~\eqref{Eq:Z_conformal_gauge}, we have used
\begin{align}
    &\Delta_{FP}[\hat{g}e^\phi]=\Delta_{FP}[\hat{g}]e^{-\frac{26}{48\pi}S_L[\hat{g};\phi]},
    &&Z_M[\hat{g}e^\phi]=Z_M[e^\phi] e^{\frac{c}{48\pi}S_L[\hat{g};\phi]}~.
\end{align}

Since $\mathcal{D}_1\phi$ is inconvenient because of $e^\phi$ in the norm \eqref{Eq:phi_norm_1},
we rewrite it in terms of the measure $\mathcal{D}_0\phi$ that is induced from the standard norm,
\begin{align}
    ||\delta\phi||^2_0=\int d^2x \sqrt{\hat{g}}\,(\delta\phi)^2~.
\end{align}
Then the parition function becomes~\cite{David:1988hj,Distler:1988jt}
\begin{align}
    Z=\int d\tau\,\mathcal{D}_0\phi\, \Delta_{FP}[\hat{g}]Z_M[\hat{g}]e^{-S[\phi;\hat{g}]}~,
\label{Eq:Z_renormalized}\end{align}
where $S[\phi;\hat{g}]$ is the (renormalized) Liouville action,
\begin{align}
    S[\phi;\hat{g}]=\frac{1}{2\pi}\int d^2x\left(\partial\phi\bar{\partial}\phi+\frac{1}{4}Q\sqrt{\hat{g}}\hat{R}\phi+\mu_1\sqrt{\hat{g}}e^{\alpha\phi}\right)~.
\label{Eq:renormalized_action}\end{align}
Here we have
\begin{align}
    &Q=\sqrt{\frac{25-c}{3}},
    &&\alpha=\frac{\sqrt{25-c}-\sqrt{1-c}}{\sqrt{12}}~.
\label{Eq:Q_and_alpha}\end{align}
Note that Eq.~(\ref{Eq:Z_renormalized}) is reduced to the semi-classical Liouville theory in the limit $c\to-\infty$.
There is another branch of $\alpha$ which does not have a sensible semi-classical limit (and usually is discarded), which we denote by $\tilde{\alpha}$:
\begin{align}
\tilde{\alpha}=\frac{\sqrt{25-c}+\sqrt{1-c}}{\sqrt{12}}.
\label{Eq:wrong_branch}\end{align}

\subsection{The absence of the Coleman mechanism}\label{Sec:absence}
In this Section, we show that the Coleman mechanism does not work for 2d Euclidean gravity coupled to conformal matter with $c\leq1$.
To do so, we compare the number of random surfaces with different topologies.

The partition function of a 2d manifold with a given topology and area $A$ is given by
\begin{align}
Z(A) = \int \frac{\mathcal{D}g_{\mu\nu}}{\text{vol}\left(\text{Diff}\right)}Z_M[g_{\mu\nu}]\delta\left(\int d^2x\sqrt{\mathrm{det}\,g_{\mu\nu}}-A\right)~,
\end{align}
which can also be viewed as the number of random surfaces of area $A$.
As reviewed in Section~\ref{Sec:DDK}, in the conformal gauge we have
\begin{align}
    Z(A)=\int d\tau\int \mathcal{D}_0\phi\, Z_M[\hat{g}]e^{-S[\hat{g};\phi]}\delta\left(\int \sqrt{\hat{g}}\,e^{\alpha\phi}d^2x-A\right)~.
\end{align}
Here we are interested in the string susceptibility $\Gamma$ defined by~\cite{Weingarten:1979gn,Eguchi:1982fe}
\begin{align}
    Z(A)\sim A^{\Gamma-3}~,
\end{align}
which is a generalization of the central limit theorem for random walks, see Appendix~\ref{Sec:Random_Walk}. 

$\Gamma$ is obtained by a scaling argument as follows~\cite{David:1988hj,Distler:1988jt} (see Ref.~\cite{Knizhnik:1988ak} for genus zero case).
By shifting $\phi$ as
\begin{align}
    \phi\to\phi+\frac{\log A}{\alpha}~,
\end{align}
the measure $\mathcal{D}_0\phi$ is invariant while the action \eqref{Eq:renormalized_action} is shifted as
\begin{align}
    S[\hat{g};\phi]\to S[\hat{g};\phi]+Q\chi\frac{\log A}{2\alpha}~.
\end{align}
Here $\chi$ is the Euler number of the 2d manifold. 
The change of the delta function is
\begin{align}
    \delta\left(\int \sqrt{\hat{g}}\,e^{\alpha\phi}d^2x-A\right)\to\frac{1}{A}\delta\left(\int \sqrt{\hat{g}}\,e^{\alpha\phi}d^2x-1\right)~.
\end{align}
Putting altogether, we obtain
\begin{align}
&Z(A) = \left.Z\right|_{A=1} A^{-\frac{\chi Q}{2\alpha}-1}= \left.Z\right|_{A=1} A^{-b\frac{\chi}{2}-1}~,
&&b=\frac{25-c+\sqrt{(1-c)(25-c)}}{12}~,
\label{Eq:scaling}\end{align}
Note that, in the region $c\leq1$ where quantum gravity is well defined, $b$ is bounded from below:
\begin{align}
&b\geq 2, 
&&\text{for \quad$c\leq1$}~.\label{ineqb}
\end{align}
For $c>1$, $Z(A)$ becomes complex which signals an instability of the spacetime against the formation of pinches \cite{Kawai:1983nq,Ambjorn:1985dn}.

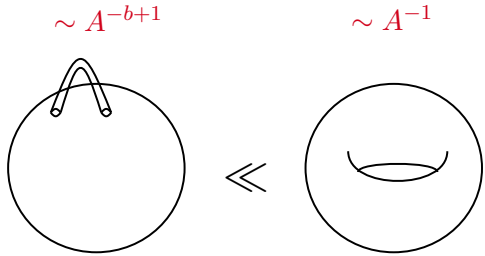
\begin{figure}
    \centering

\tikzset{every picture/.style={line width=0.75pt}} 

\begin{tikzpicture}[x=0.75pt,y=0.75pt,yscale=-1,xscale=1]

\draw  
(28.5,53.5) .. controls (28.5,30.03) and (48.43,11) .. (73,11) .. controls (97.58,11) and (117.5,30.03) .. (117.5,53.5) .. controls (117.5,76.97) and (97.58,96) .. (73,96) .. controls (48.43,96) and (28.5,76.97) .. (28.5,53.5) -- cycle ;
\draw 
(50,25.5) .. controls (64,-10.56) and (66,-10.69) .. (80.5,25.5) ;
\draw 
(55,25.5) .. controls (64,-4.56) and (66,-4.69) .. (75.5,25.5) ;
\draw  
(50,25.5) .. controls (52,23) and (53,23) .. (55,25.5) ;
\draw  
(50,25.5) .. controls (52,28) and (53,28) .. (55,25.5) ;
\draw  
(75.5,25.5) .. controls (77.5,23) and (78.5,23) .. (80.5,25.5) ;
\draw  
(75.5,25.5) .. controls (77.5,28) and (78.5,28) .. (80.5,25.5) ;

\draw  
(178.5,53.5) .. controls (178.5,30.03) and (198.43,11) .. (223,11) .. controls (247.58,11) and (267.5,30.03) .. (267.5,53.5) .. controls (267.5,76.97) and (247.58,96) .. (223,96) .. controls (198.43,96) and (178.5,76.97) .. (178.5,53.5) -- cycle ;
\draw 
(200,45) .. controls (200,65) and (250,65) .. (250,45) ;
\draw 
(205,55.5) .. controls (205,50) and (245,50) .. (245,55.5) ;

\draw (135.5,50) node [anchor=north west][inner sep=0.75pt]  [rotate=0] [align=left] {{\fontsize{18pt}{18pt}\selectfont $\ll$}};
 \draw (200,-30) node [anchor=north west][inner sep=0.75pt]    {$\textcolor[rgb]{0.82,0.01,0.11}{\sim A^{-1}}$};
 \draw (50,-30) node [anchor=north west][inner sep=0.75pt]    {$\textcolor[rgb]{0.82,0.01,0.11}{\sim A^{-b+1}}$};

\end{tikzpicture}

    \caption{For fixed large area $A$, the number of random tori is much larger than the number of random spheres with a thin tube.}
    \label{Fig:random_surface}
\end{figure}

This can be used to discuss the magnitude of quantum fluctuations of spacetime due to baby universes in 2d Euclidean gravity. As an example, consider a situation in which spacetime is a macroscopic 2d sphere, and a tiny tubular wormhole is attached to it (see the left figure of Fig.~\ref{Fig:random_surface}). This represents the process of a tiny circular baby universe branching off from the circular mother universe and being absorbed back into the mother universe. Overall, the spacetime is a 2d torus. On the other hand, if the spacetime is a macroscopic 2d torus, it represents the process of a circular mother universe splitting into two macroscopic mother universes, which then merge back into a single mother universe (see the right figure of Fig.~\ref{Fig:random_surface}). If the former is significantly non-zero compared to the latter, the former can be regarded as a quantum correction due to microscopic fluctuations (baby universes) to the 2d spherical spacetime. 
Then after integrating out the contributions of small wormholes, we obtain an effective field theory in which the cosmological constant appears as a dynamical parameter~\cite{Coleman:1988tj,Klebanov:1988eh}.
Unfortunately, as we will see below, this is not the case in the simple 2d Euclidean gravity.

From Eq.~\eqref{Eq:scaling} with $\chi=0$, we obtain
\begin{align}
\left(\text{\# of random surfaces with area $A$ and topology $T^2$}\right) \sim A^{-1}~.
\label{Eq:all_contribution}\end{align}
This equation includes both the contribution of a macroscopic sphere with a microscopic wormhole and the contribution of a macroscopic torus. The former is estimated as follows.
\begin{align}
\left(\text{\# of random spheres with a microscopic wormhole}\right) \sim A^{-b-1}\cdot A^2 = A^{-b+1}~.
\label{Eq:small_contribution}\end{align}
Here $A^{-b-1}$ is the number of random spheres, and $A^2$ is the number of ways to attach the endpoints of the microscopic wormhole.
From Eq.~(\ref{ineqb}), we see that $A^{-b+1}\ll A^{-1}$ for large $A$ and that the effect of the microscopic wormhole is negligibly small compared to the macroscopic topology change.
In other words, there is no special mechanism to enhance the effect of small wormholes.
This is due to the fact that the gravitational coupling is dimensionless in two dimensions ($\sim 1/\sqrt{c}$), and consequently, there is no intrinsic difference between small and large wormholes~\footnote{This is also pointed out in e.g. \cite{Polchinski:1989fn}.}.

\subsection{Modification of the model}
The Coleman mechanism may be achieved by modifying the model. To do so, we start with the large-$N$ limit of a matrix model.
\begin{align}
S= N\left( \frac{1}{2}\mathrm{tr}\phi^2- \frac{\lambda}{3}\mathrm{tr}\phi^3 \right),
\label{Eq:modifeid_matrix_model}\end{align}
where $\phi$ is an $N\times N$ Hermite matrix.
This describes the 2d pure gravity ($c=0$) in the scaling limit, see Ref.~\cite{Francesco_1995} for a review and the references therein. 

One of the possible modifications is to consider the action as defined as a polynomial of the local actions~\cite{Hamada:2015dja}. 
Here we consider the simplest modification, namely adding a term corresponding to $\left(\int d^2x\sqrt{g}\right)^2$:
\begin{align}
S&= N\left( \frac{1}{2}\mathrm{tr}\phi^2- \frac{\lambda_0}{3}\mathrm{tr}\phi^3 \right) - \frac{1}{2}C \left(\frac{1}{3}\mathrm{tr}\phi^3\right)^2~. 
\label{Eq:modified_action}\end{align}
In fact, in terms of Feynman diagrams, the last term represents the insertion of a pair of $\phi^3$ vertices, which is just a discretization of $\left(\int d^2x\sqrt{g}\right)^2$. 
Then the partition function is formally evaluated as
\begin{align}
Z&= \int d\phi \exp\left( -N\left( \frac{1}{2}\mathrm{tr}\phi^2 - \frac{\lambda_0}{3}\mathrm{tr}\phi^3 \right) + \frac{1}{2}C \left(\frac{1}{3}\mathrm{tr}\phi^3\right)^2\right)
\nonumber\\&=\int d\lambda \int d\phi \exp\left( -N\left( \frac{1}{2}\mathrm{tr}\phi^2 - \frac{\lambda+\lambda_0}{3}\mathrm{tr}\phi^3 \right) - \frac{N^2}{2C}\lambda^2\right)
\nonumber\\&=\int d\lambda \,Z_{\phi^3}(\lambda+\lambda_0) \exp\left(-\frac{N^2}{2C}\lambda^2\right)
\nonumber\\&=\int d\lambda \,Z_{\phi^3}(\lambda) \exp\left(-\frac{N^2}{2C}(\lambda-\lambda_0)^2\right)~,
\label{Eq:modified_Z}\end{align}
where
\begin{align}
Z_{\phi^3}(\lambda)=\exp\left(Z_{\text{single}}(\lambda)\right):= \int d\phi \exp\left(-N\left(\frac{1}{2}\mathrm{tr}\phi^2- \frac{\lambda}{3}\mathrm{tr}\phi^3 \right)\right)~.
\end{align}
This fulfills Coleman's idea of considering ensembles of various coupling constants simultaneously:
\begin{align}
Z=\int d\lambda \,\exp\left(-\frac{N^2}{2C}(\lambda-\lambda_0)^2\right)\exp\left(Z_{\text{single}}(\lambda)\right)
=:\int d\lambda w(\lambda) \exp\left(Z_{\text{single}}(\lambda)\right)~.
\label{Eq:average}\end{align}
Then $\lambda$ is fixed to the peak of the integrand.

Unfortunately, however, 
the formal partition function \eqref{Eq:modified_Z} is divergent. 
This can be seen in both views: the sum over the area and the integration over the cosmological constant.

We start with the former view. In terms of Feynman diagrams, $A$ is the number of vertices. A single insertion of $C(\mathrm{Tr}(\phi^3))^2$ can be viewed as the insertion of two separate vertices, giving rise to a factor,
\begin{align}
    kN^2C\times A^2~,
\end{align}
where $k$ is a positive $\mathcal{O}(1)$ constant.
By adding multiple insertions, we obtain the factor
\begin{align}
\exp\left(kN^2C\times A^2\right)~.
\end{align}
This indicates that the sum over $A$ does not converge because the partition function (the number of planar Feynman diagrams) when $C=0$ is bounded by an exponential function of $A$.

In the latter view, the partition function of a single universe $Z_\text{single}(\lambda)$ is not well defined when $\lambda$ is above a critical value.
In the matrix model, $Z_\text{single}$ is given by a power series of $\lambda$ with positive coefficients,
\begin{align}
Z_\text{single}(\lambda) = N^2\left(a_0+a_1\lambda + a_2\lambda^2 +\cdots \right)= N^2\sum_{A=0}^{\infty} a_A \lambda^A~.
\end{align}
It has a convergence radius $\lambda_c$, and the critical behavior near $\lambda_c$ is given by\footnote{$\log\lambda$ and $(\lambda_c-\lambda)/\lambda_c$ are regarded as the bare and renormalized cosmological constant, respectively.}
\begin{align}
\sim \mathrm{const.}N^2\left(\frac{\lambda_c-\lambda}{\lambda_c}\right)^{b}~.
\end{align}
The complete partition function is obtained by substituting $Z_\text{single}(\lambda)$ into Eq.~\eqref{Eq:average}, but it is divergent because the integration over $\lambda$ includes the region $\lambda>\lambda_c$.

On the other hand, even though the formal partition function \eqref{Eq:modified_Z} is divergent, the large $N$ limit of the matrix model itself~\eqref{Eq:modifeid_matrix_model} is known to be well-defined.
Indeed, the interactions of the form $({\rm tr}(\phi^n))^2$ are called as the touching interactions~\cite{Das:1989fq,Durhuus:1994tu,Klebanov:1994pv,Barbon:1995dx,Ambjorn:2016lkl} because they glue two isolated surfaces at a point.\footnote{The models with different $n$ are expected to be in the same universality class.}   
In particular, in the large $N$, it was shown that the interaction $({\rm tr}(\phi^2))^2$ changes scaling of $Z(A)$ from Eq.~\eqref{Eq:scaling} to\footnote{The model~\eqref{Eq:modified_action} is for $c=0$, but the analysis can be extended to $0\leq c<1$.}
\begin{align}
&Z(A) = \left.Z\right|_{A=1} A^{-\tilde{b}\frac{\chi}{2}-1},
&&\tilde{b}=\frac{25-c-\sqrt{(1-c)(25-c)}}{12}=-\frac{b}{1-b},
\end{align}
when the strength of the touching interaction is tuned to be a special value. 
The value of $\tilde{b}$ corresponds to the unusual branch~\eqref{Eq:wrong_branch}, and $\tilde{b}<2$ for $0\leq c<1$.
Then, the counting of the random surface in Fig.~\ref{Fig:random_surface} is replaced by
\begin{align}
 A^{-\tilde{b}+1}\gg A^{-1},
\label{Eq:modified_counting}\end{align} 
which indicates that the contribution from the microscopic wormhole is significant.
It is an open question, in spite of this fact, why the interpretation in terms of the ensemble average~\eqref{Eq:modified_Z} does not work.
We leave the investigation of this interesting question to future publications.
Instead of that, in this paper, we explore Lorentzian gravity as an alternative possibility to realize the Coleman mechanism.

Therefore, we consider a Lorenz model such as
\begin{align}
Z_L:= \int d\lambda  \exp\left(i N^2\frac{(\lambda-\lambda_0)^2}{2C}\right) \exp\left(Z_{\text{single}}(\lambda)\right)
\end{align}
instead of the Euclidean model.
The remainder of this paper will explore this possibility.

\begin{figure}
    \centering

\tikzset{every picture/.style={line width=0.75pt}} 

\begin{tikzpicture}[x=0.75pt,y=0.75pt,yscale=-1,xscale=1]

\draw  
(195,96) .. controls (198.43,96) and (185.5,76) .. (183,53.5) .. controls (182,46) and (182,38) .. (183,30.5) .. controls (187,10) and (191.33,5) .. (195,2);
\draw 
(260,96) .. controls (260.58,96) and (263,64.97) .. (263.5,58.5) .. controls (264,50) and  (264,41.5) .. (262.5,33.5) .. controls (260,20) and (250.33,10) .. (240,2);
\draw 
[shift={(-10,0)}] (225.5,25) .. controls (235,20) and (235,70) .. (225.5,65) ;
\draw 
[shift={(-10,0)}] (225.5,25) .. controls (220,25) and (220,65) .. (225.5,65) ;
\draw [color={rgb, 255:red, 0; green, 38; blue, 209 }  ,draw opacity=1 ][line width=1]    (240,2) .. controls (230,-5) and (205,-5) .. (195,2) .. controls (205,9) and (230,9) .. (240,2);
\draw [color={rgb, 255:red, 0; green, 38; blue, 209 }  ,draw opacity=1 ][line width=1]    (260,96) .. controls (238.2,86) and (216.6,86) .. (196,96) .. controls (216.6,106) and (238.2,106) .. (260,96);

 \draw (183,38) node [anchor=north west][inner sep=0.75pt]    {\textcolor[rgb]{0.82,0.01,0.11}{{\fontsize{12pt}{12pt}\selectfont $t_A$}}};
  \draw (240,38) node [anchor=north west][inner sep=0.75pt]    {\textcolor[rgb]{0.82,0.01,0.11}{{\fontsize{12pt}{12pt}\selectfont $t_B$}}};
    \draw (130,48) node [anchor=north west][inner sep=0.75pt] [rotate=-270]   {\textcolor[rgb]{0,0,0}{{\fontsize{12pt}{12pt}\selectfont time}}};

\draw [shift={(10,0)}, rotate = 0] [color={rgb, 255:red, 208; green, 2; blue, 27 }  ,draw opacity=1 ]   (227,65) -- (227,24) ;
\draw [shift={(237,20)}, rotate = 90] [color={rgb, 255:red, 208; green, 2; blue, 27 }  ,draw opacity=1 ][line width=0.75]    (10.93,-3.29) .. controls (6.95,-1.4) and (3.31,-0.3) .. (0,0) .. controls (3.31,0.3) and (6.95,1.4) .. (10.93,3.29)   ;
\draw [shift={(237,68)}, rotate = -90] [color={rgb, 255:red, 208; green, 2; blue, 27 }  ,draw opacity=1 ][line width=0.75]    (10.93,-3.29) .. controls (6.95,-1.4) and (3.31,-0.3) .. (0,0) .. controls (3.31,0.3) and (6.95,1.4) .. (10.93,3.29);
\draw [shift={(-25,0)}, rotate = 0] [color={rgb, 255:red, 208; green, 2; blue, 27 }  ,draw opacity=1 ]   (227,65) -- (227,24) ;
\draw [shift={(202,20)}, rotate = 90] [color={rgb, 255:red, 208; green, 2; blue, 27 }  ,draw opacity=1 ][line width=0.75]    (10.93,-3.29) .. controls (6.95,-1.4) and (3.31,-0.3) .. (0,0) .. controls (3.31,0.3) and (6.95,1.4) .. (10.93,3.29)   ;
\draw [shift={(202,68)}, rotate = -90] [color={rgb, 255:red, 208; green, 2; blue, 27 }  ,draw opacity=1 ][line width=0.75]    (10.93,-3.29) .. controls (6.95,-1.4) and (3.31,-0.3) .. (0,0) .. controls (3.31,0.3) and (6.95,1.4) .. (10.93,3.29);

\draw [shift={(-75,-20)}, rotate = 0]  (227,115) -- (227,24) ;
\draw [shift={(152,0)}, rotate = 90] [line width=0.75]    (10.93,-3.29) .. controls (6.95,-1.4) and (3.31,-0.3) .. (0,0) .. controls (3.31,0.3) and (6.95,1.4) .. (10.93,3.29)   ;

\end{tikzpicture}

    \caption{The topology change of the universe. The process in which one universe splits into two and then merges into one again is shown.}
    \label{Fig:splitting}
\end{figure}
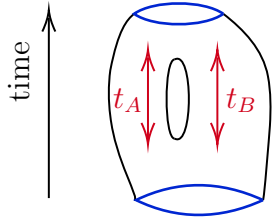

Before concluding this section, a general point should be made about Lorentzian gravity.
From Geroch's theorem~\cite{Geroch:1967fs}, a singularity exists when a topology change occurs.
This creates an ambiguity about time in the separated universes.
For example, consider the process of one universe splitting into two and merging back into one.
In this case, the relationship between the time of each separated universe is not a priori clear ($t_A$ may or may not be equal to $t_B$ in Figure \ref{Fig:splitting}).
This is in contrast to the case of Euclidean gravity, where a common proper time ($t_A=t_B$) should be chosen.

\section{Coleman mechanism in Lorentzian multiverse}\label{Sec:Lorentzian}
In this section, we propose a non-Hermitian Hamiltonian describing a Lorentzian multiverse and show that the Coleman mechanism actually works in such a model. 
We also discuss various phenomenological implications. 
 
\subsection{Lorentzian Model}
We start with the Hamiltonian of the universe in the mini-superspace approximation of 4-dimensional Lorentzian gravity.
\aln{
H_{\rm mini}^{}=-\frac{p_a^2}{2M_{\rm Pl}^2a}+\frac{a^3}{6}\lambda(a)~,\quad p_a^{}=M_{\rm Pl}^2a\dot{a}~,\label{mini-superspace Hamiltonian}
}
where $\lambda(a)$ is the energy density of the universe.\footnote{One can see that $H_{\rm mini}^{}=0$ corresponds to the Freedman equation. 
} 
It is convenient to choose the volume $l=a^3/3$ as a dynamical variable instead of $a$.  
Eq.~(\ref{mini-superspace Hamiltonian}) is then rewritten as 
\aln{
H_{\rm mini}^{}=-l\frac{p_l^2}{2M_{\rm Pl}^2}+\frac{l}{2}\lambda(l)
~,\quad p_l^{}=\frac{M_{\rm Pl}^2\dot{l}}{l}~. 
}
In the following, we set $M_{\rm Pl}^{}=1$.  
To describe the multiverse including baby universes, we consider the following second quantized Hamiltonian.
\aln{
\hat{H}_{}^{}&=\frac{1}{2}\int_0^\infty dl\hat{\psi}^\dagger(l)\left(\hbar^2\frac{d}{dl}l\frac{d}{dl}+l\lambda(l)+c\,l(\hat{a}+\hat{a}^\dagger)\right)\hat{\psi}(l)
+\int_0^\infty dll\rho^*(l)\psi^\dagger(l)~,
}
where we impose the commutation relations
\begin{align}
    [\psi(l),\psi^\dagger (l')]=\delta(l-l')~,\quad [\hat{a},\hat{a}^\dagger]=1~. 
\end{align}
Here $\hat{a}$ and $\hat{a}^\dagger$ are the annihilation and creation operators of the baby universe, and $\hat{\psi}(l)$ and $\hat{\psi}^\dagger(l)$ are the annihilation and creation operators of the mother universe of length $l$. 
Also $\hat{\psi}^\dagger(l)(\hat{a}+\hat{a}^\dagger)\hat{\psi}(l)$ represents the emission and absorption of the baby universe from the mother universe. 
We note that $\hat{q}=\hat{a}+\hat{a}^\dagger$ is a conserved quantity because it commutes with the Hamiltonian.  

We have not introduced terms such as the first and second terms on the right hand side of Eq.(\ref{Eq:Euclidean_Hamiltonian}) representing the splitting and merging of the mother universes. This is because in Lorentzian gravity, topology changes of the universe occur through tunneling, but the Euclidean action representing the tunneling barrier between the 3-dimensional macroscopic universes is macroscopically large, so the tunneling probability is strongly suppressed.

\begin{figure}
\begin{center}
\includegraphics[width=6cm]{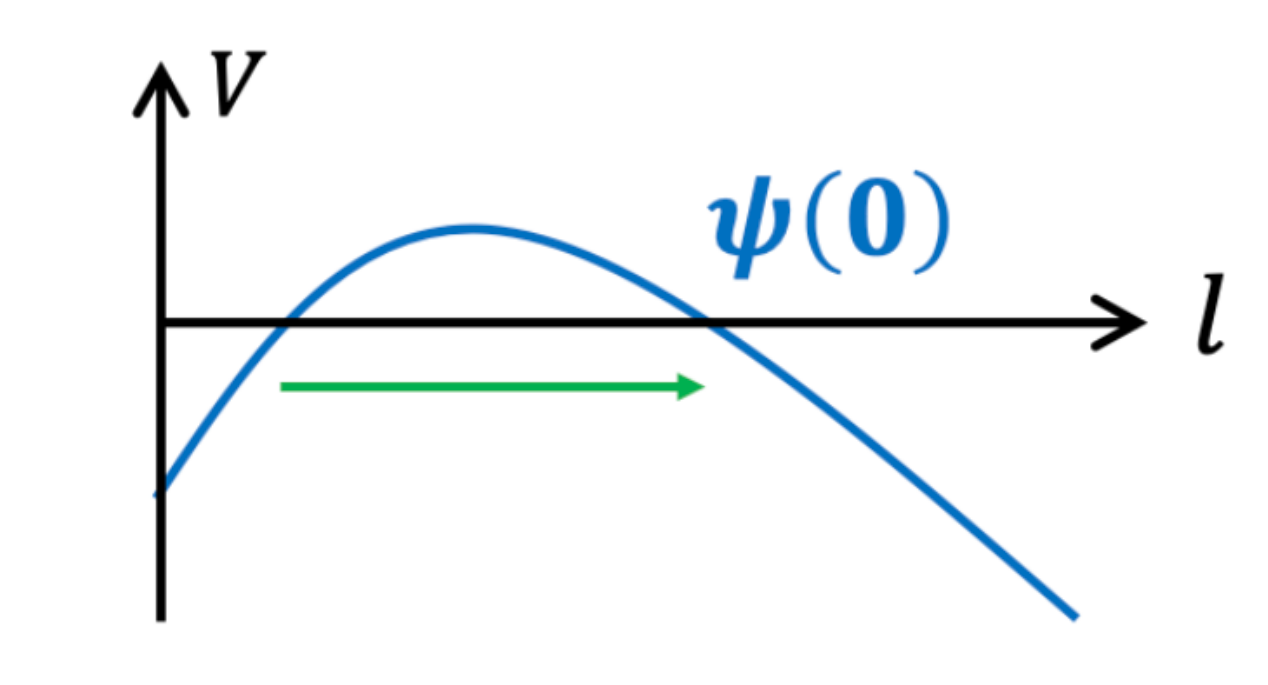}
\end{center}
\caption{Creation of a small mother universe from nothing.  
}
\label{fig:tunneling}
\end{figure}

The last term in $\hat{H}_{}^{}$ describes the process by which a small mother universe arises from nothing by the tunneling effect~\cite{Vilenkin:1982de} (see Fig.~\ref{fig:tunneling}). 
It is important to note that there is a term that creates the mother universe, but not a term that annihilates the mother universe.
This is because if the matter field of the universe is highly excited, the overlap with the ground state is so small that the probability of the universe becoming nothing after a big crunch is expected to be almost zero.\footnote{
Such a contracting universe may bounce back and repeat the cycle due to some quantum gravity effects.
}
For this reason, it is natural to consider the non-hermitian effective Hamiltonian of the multiverse. This is important in the discussion that follows.

A general initial state is given by
\aln{
|i\rangle=\int_{-\infty}^\infty dq f(q)|q\rangle \otimes |i,q\rangle_M^{}~,  
}
where $|q\rangle$ and $|i,q\rangle_M^{}$ are the states of the baby universe and the mother multiverse, respectively.
Then, the time evolution is given by
\aln{e^{-i\hat{H}t/\hbar}|i\rangle&=\int_{-\infty}^\infty dqf(q) |q\rangle \otimes e^{-i\hat{H}_{q}^{}t/\hbar}|i,q\rangle_M^{} ~,
\label{rhoM}
} 
where  
\aln{
\hat{H}_{q}^{}&=\frac{1}{2}\int_0^\infty dl\hat{\psi}^\dagger(l)\left(\hbar^2\frac{d}{dl}l\frac{d}{dl}+l\lambda_{\rm eff}^{}(l)\right)\hat{\psi}(l)
+\int_0^\infty dll\rho^*(l)\psi^\dagger(l)~,
\label{q hamiltonian}
\\
\lambda_{\rm eff}^{}(l)&=\lambda(l)+c\,q~. 
}
It is clear that $c\,q$ plays the role of the cosmological constant.
In the following, we denote the constant piece of $\lambda_{\rm eff}^{}(l)$ as $\lambda$. 
Then the integral of $q$ in Eq.~(\ref{rhoM}) can be regarded as an ensemble average over the cosmological constants.    
Note that the generalization to other coupling constants is straightforward.  
In fact, if we want to realize an ensemble average over the coupling constant $g_i^{}$  corresponding to a local operator ${\cal O}_i^{}(x)$, we can simply introduce another baby universe via 
\aln{
(\hat{a}_i^{}+\hat{a}_i^\dagger)\int d^dx \,{\cal O}_i^{}(x)~,
}
where $\hat{q}_i^{}=\hat{a}_i^{}+\hat{a}_i^\dagger$ plays the role of $g_i^{}$. 
This has already been discussed in the original work~\cite{Coleman:1988tj,Coleman:1988cy}.

From Eq.~(\ref{rhoM}), we see that the quantum state of the mother multiverse with cosmological constant $\lambda$ is given by\footnote{For simplicity, we do not write subscript $M$, but $|\Psi_\lambda^{}(t)\rangle$ is the state of the mother multiverse.}
\begin{align}
    \langle \lambda|e^{-i\hat{H}t/\hbar}|i\rangle=f(\lambda)e^{-i\hat{H}_\lambda^{}t/\hbar}|i,\lambda\rangle_M^{}=:f(\lambda)|\Psi_\lambda^{}(t)\rangle~.
\label{Eq:psi_q}\end{align}
In general, for non-unitary systems, there is no clear definition of probability a priori. 
However, it is natural to assume that the probability of observing that the cosmological constant is $\lambda$ is proportional to the norm of Eq.~\eqref{Eq:psi_q},
i.e.
\aln{
P(t,\lambda)\propto|f(\lambda)|^2\langle \Psi_\lambda^{}(t)| \Psi_\lambda^{}(t)\rangle~. 
\label{P}
}
If the time evolution is unitary, this reduces to $|f(\lambda)|^2$. 
Then, the cosmological constant is simply determined by the initial wavefunction $f(\lambda)$.    
On the other hand, in the non-unitary model we are considering, the behavior of Eq.~(\ref{P}) is non-trivial.   
The Heisenberg equation of $\hat{\psi}(l)$ is
\aln{
i\hbar \frac{\partial \hat{\psi}}{\partial t}&=[\hat{\psi},\hat{H}_q]
=\frac{1}{2}\left(\hbar^2\frac{d}{dl}l\frac{d}{dl}+l\lambda_{\rm eff}^{}(l)\right)\hat{\psi}+l\rho^*(l)~,\label{Heisenberg eq}
}
which looks like a Schr\"{o}dinger equation with a source term $\rho^*(l)$. 
The inhomogeneous terms in this equation imply that universes of various sizes are generated per unit time. On the other hand, the homogeneous terms mean that the generated universe expands toward infinity.\footnote{One can interpret $-\lambda_{\rm eff}^{}(l)$ as the potential energy for the size of the universe.}  
Therefore, it is expected that the system reaches a stationary coherent state $|\Psi_{st}^{}\rangle$ after a sufficiently long time, so that $\psi_{st}^{}(l):=\langle\Psi_{st}|\hat{\psi}(l)|\Psi_{st}^{}\rangle$ satisfies 
\aln{
&\left(\hbar^2\frac{d}{dl}l\frac{d}{dl}+l\lambda_{\rm eff}^{}(l)\right)\psi_{st}^{}(l)+l\rho^*(l)=0~.
\label{eq:equilibrium}
}
In fact, we can show that the following $|\Psi_{st}^{}\rangle$ satisfies $\hat{H}_\lambda^{}|\Psi_{st}^{}\rangle=0$ if Eq.~(\ref{eq:equilibrium}) is satisfied: 
\aln{
|\Psi_{st}^{}\rangle={\cal N}^{1/2}\exp\left(\int_0^\infty dl\hat{\psi}^{\dagger}(l)\psi_{st}^{}(l)\right)|0\rangle~,
}
where ${\cal N}$ is the normalization constant.
Note that this stationary multiverse state corresponds to the multiverse partition function in the path-integral formulation~\cite{Coleman:1988tj,Coleman:1988cy,Kawai:2011rj,Kawai:2011qb,Hamada:2015dja}. In the present formulation, such a state emerges naturally from the non-hermitian many-body Hamiltonian.

The probability distribution of the cosmological constant is now given by
\aln{
P_{st}^{}(\lambda)=|f(\lambda)|^{2}\langle \Psi_{st}|\Psi_{st}^{}\rangle 
=|f(\lambda)|^{2}{\cal N}
 \exp\left(\frac{1}{2}\int_0^\infty dl|\psi_{st}^{}(l)|^2\right)~,\label{equilibrium distribution}
} 
As argued above, the stationary state is expected as a result of the balance between inhomogeneous and homogenous terms. Here we have implicitly assumed that generated universe expands toward infinity.
This is equivalent to the assumption that we consider an ensemble average of coupling constants where $\lambda_{\rm eff}^{}(l)\geq0$ is satisfied.\footnote{This may be viewed as an anthropic principle in a weak sense. We require that the universe can be large, though we do not require the formation of the galaxy~\cite{Weinberg:2000yb}.
}
In the following, we consider a source term of the form
\aln{
l\rho^*(l)=\nu\epsilon \delta(l-\epsilon)~,
}
which means that a mother universe of initial size $\epsilon$ is generated at a rate of $\nu$ per unit time. 
Moreover, we simply put $\nu\epsilon\rightarrow \nu$.

We now summarize the WKB solution for $\psi_{st}^{}(l)$.  
By expanding $\psi_{st}^{}(l)$ as 
\aln{\psi_{st}^{}(l)=\exp\left(\frac{i}{\hbar}S_0^{}+iS_1^{}+\cdots\right)~,
}
we have
\aln{
{\cal O}(\hbar^0)&:\ \left(\frac{dS_0^{}}{dl}\right)^2=\lambda_{\rm eff}^{}(l)~,
\\
{\cal O}(\hbar^1)&:\ -2l\frac{dS_0^{}}{dl}\frac{dS_1^{}}{dl}+i\left(\frac{dS_0^{}}{dl}+l\frac{d^2S_0^{}}{dl^2}\right)=0~.
}
These are solved as
\aln{S_0^{}(l)=\pm \int^l dl' \sqrt{\lambda_{\rm eff}^{}(l')}~,\quad S_1^{}(l)=\frac{i}{2}\left(\log \sqrt{\lambda_{\rm eff}^{}(l)}+\log l\right)~.
}
Thus, the general WKB solution for $l>\epsilon$ is given by
\aln{
\psi_{st}^{\rm WKB}(l)=\frac{A}{\sqrt{l}\,\lambda_{\rm eff}^{}(l)^{1/4}} e^{\frac{i}{\hbar}\int^l dl'\sqrt{\lambda_{\rm eff}^{}(l')}}+\frac{B}{\sqrt{l}\,\lambda_{\rm eff}^{}(l)^{1/4}}e^{-\frac{i}{\hbar}\int^l dl'\sqrt{\lambda_{\rm eff}^{}(l')}}~, 
\label{WKB solution}
}
where $A$ and $B$ are constants proportional to $\nu$ and are independent of $l$. 
In the following, we put $\hbar=1$. 
By substituting this into Eq.~\eqref{equilibrium distribution}, we obtain
\aln{
P_{st}^{}(\lambda,\{g_i^{}\})
&={\cal N}|f(\lambda,\{g_i^{}\})|^2
 \exp\left[\frac{1}{2}\int_0^{l_{\rm IR}^{}}\frac{d\log l}{\lambda_{\rm eff}^{}(l)^{1/2}}
\left|A+B e^{-2i\int^l dl'\sqrt{\lambda_{\rm eff}^{}(l')}}\right|^2
\right]~, 
\label{equilibrium distribution 2}
}
where an IR cutoff $l_{\rm IR}$ as a maximum size of the universe is introduced, and we have added the other coupling constants, $\{g_i^{}\}$, to the argument of $P_{st}$ to emphasize that $\lambda_{\rm eff}$ may have a dependence on $g_i^{}$.

Since the integrand of this exponent is clearly nonnegative, the dominant contribution to the integral will come from the neighborhood of the minimum of $\lambda_{\rm eff}$, unless the factor with absolute value happens to be small.
In the remainder of this section, we will use Eq.~\eqref{equilibrium distribution 2} to compute the probability distributions of the cosmological constant and the other coupling constants under these assumptions. 
Here we consider the following two cases:
\begin{itemize}
\item Spatially flat universe with a cosmological constant $\lambda$ 
(Section~\ref{Sec:fine-tuning}).
\item Universe with the positive spatial curvature, a cosmological constant $\lambda$, and a radiation or matter component of energy density (Section~\ref{Sec:MEP}).
\end{itemize}
In the former, the cosmological constant is taken into account. In the latter, not only that, but also the coupling constants that affect the energy density are considered. 
By explicitly computing Eq.~\eqref{equilibrium distribution 2}, we show that in both cases the probability distribution of $\lambda$ has a sharp peak near zero.
This implies that the cosmological constant is fine-tuned to zero.
Furthermore, in the latter case, the probability of $g_i$ is found to be maximum at the point where the energy density of radiation or matter in the late universe is maximum. 
We call this the maximum entropy principle~\cite{Kawai:2011rj,Kawai:2011qb,Hamada:2014ofa,Hamada:2014xra,Hamada:2015dja}, or the maximum matter principle.

\subsection{Fine-tuning of the cosmological constant}\label{Sec:fine-tuning}
We consider the spatially flat universe with the cosmological constant $\lambda$. 
The effective vacuum energy is just a constant:
\begin{align}
\lambda_{\rm eff}=\lambda
\end{align}
As discussed below Eq.~\eqref{equilibrium distribution}, we assume  $\lambda$ is non-negative.

The probability distribution Eq.~(\ref{equilibrium distribution 2}) becomes
\aln{
P_{st}^{}(\lambda)
&={\cal N}
 |f(\lambda)|^2\exp\left[\frac{1}{2}\int_0^{l_{\rm IR}^{}}\frac{d\log l}{\lambda^{1/2}}\left|A+B e^{-2i\int^l dl'\sqrt{\lambda}}\right|^2
\right]~. 
\nonumber\\
&\sim|f(\lambda)|^2\exp\left(\frac{\log l_{\rm IR}^{}}{\lambda^{1/2}}\right)~,
\label{IR distribution}
}
which has a sharp peak at $\lambda=0$.
Note that the constants $A$ and $B$ are not important as these are independent of $l$.
Here, we again impose the IR cutoff $l_{\rm IR}$ as a maximum size of the universe and assumed that the initial wavefunction $f(\lambda)$ does not have strong parameter dependences compared to the singular exponential factor. 
Therefore, the cosmological constant is fixed to be zero. 
%

\subsection{Maximum entropy principle and maximum matter principle}\label{Sec:MEP}
Next, let us consider that the universe has a three-dimensional spherical topology in space and that the energy density consists of the cosmological constant and the matter or radiation component.       
The energy density of the universe is
\begin{align}
\lambda_{\rm eff}^{}(l)=
\begin{cases}
\lambda+\dfrac{S}{l^{4/3}}-\dfrac{1}{Gl^{2/3}}\quad(\text{radiation})\\ \\
\lambda+\dfrac{M}{l}-\dfrac{1}{Gl^{2/3}}\quad(\text{matter})
\end{cases},
\end{align}
where the first and second lines correspond to the radiation-dominated universe and matter-dominated universe, respectively.
For both cases, the first term is the cosmological constant and the last term is the contribution from the spatial curvature of the universe where $8\pi G=M_{\rm Pl}^{-2}=1$. 
As for the second term, $S/l^{4/3}$ and $M/l$ are the radiation and matter energy densities with $S$ and $M$ being the total entropy of the radiation and total energy of the matter, respectively.\footnote{Strictly speaking, this definition of ``entropy" is different from the usual definition of radiation entropy $S_{\rm rad}^{}\sim \rho_{\rm rad}^{3/4}a^{3}\propto T^3a^{3}$.  
In our case, we have $S\sim \rho_{\rm rad}a^4 \propto S_{\rm rad}^{4/3}$. 
}

As discussed below Eq.~\eqref{equilibrium distribution}, we assume  $\lambda_\mathrm{min}\geq0$.
As we will see explicitly,  
the dominant contribution to the integral in the probability distribution~\eqref{equilibrium distribution 2} comes from the region around the minimum of $\lambda_{\rm eff}^{}$.
To this end, we expand $\lambda_{\rm eff}^{}(l)$ around its minimum at $l=l_\mathrm{min}$ as
\begin{align}
\lambda_{\rm eff}^{}(l) = \lambda_\mathrm{min} + \lambda_\mathrm{min}^{(2)}(l-l_\mathrm{min})^2+\mathcal{O}((l-l_\mathrm{min})^3)~,
\label{Eq:expansion}\end{align}
where
\begin{align}
&\lambda_\mathrm{min}=
\begin{cases}
\lambda-\dfrac{1}{4SG^2} \\ \\
\lambda-\dfrac{4}{27G^3M^2}
\end{cases},
&&
\lambda_\mathrm{min}^{(2)}=
\begin{cases}
\dfrac{1}{72G^5S^4} \quad(\text{radiation})\\ \\
\dfrac{256}{59049G^9M^8}\quad(\text{matter})
\end{cases},
\end{align}
and
\begin{align}
l_\mathrm{min}=
\begin{cases}
(2GS)^{3/2} \quad (\text{radiation})\\
\left(3GM/2\right)^{3} \quad (\text{matter})
\end{cases}.
\end{align}
Then, we divide the integral in Eq.~\eqref{equilibrium distribution 2} into two parts.
The region around $l=l_\mathrm{min}$ and the other contribution. The former is evaluated by substituting Eq.~\eqref{Eq:expansion} into the integral: 
 \begin{align}
P_{st}^{}(\lambda,\{g_i^{}\}) &\sim |f(\lambda,\{g_i^{}\})|^2\exp\left( \int^{2l_{\mathrm{min}}}_{l_{\mathrm{min}}/2} \frac{dl}{l_\mathrm{min}}\frac{1}{\sqrt{\lambda_\mathrm{min}+\lambda_\mathrm{min}^{(2)}\left(l_\mathrm{min}-l\right)^2}}
+ \text{(other)} \right)
\nonumber\\
&\sim |f(\lambda,\{g_i^{}\})|^2\exp\left[ \frac{1}{l_{\mathrm{min}} \sqrt{\lambda_\mathrm{min}^{(2)}}} \log\left(\frac{l_\mathrm{min}^2\lambda_\mathrm{min}^{(2)}}{\lambda_\mathrm{min}}\right)+ \text{(other)}\right].
\nonumber\\&\sim
\begin{cases}
\exp\left[ G \sqrt{S} \log\left(\dfrac{1}{\lambda_\mathrm{min}}\right) + \text{(other)} \right] \quad(\text{radiation})\\ \\
\exp\left[ G^{3/2} M \log\left(\dfrac{1}{\lambda_\mathrm{min}}\right) + \text{(other)} \right] \quad(\text{matter})
\end{cases},
\label{Eq:divergent_contribution}\end{align}
where $A$ and $B$ are omitted as in Eq.~\eqref{IR distribution}.
We observe that, when $\lambda_\mathrm{min}=0$, the integral around $l=l_\mathrm{min}$ is divergent.\footnote{Note that the above divergence is interpreted as the life-time of the universe in the path-integral formulation~\cite{Kawai:2011rj,Kawai:2011qb,Hamada:2014ofa,Hamada:2014xra,Hamada:2015dja}. }
On the other hand, the other contributions to the integral are finite once the IR cutoff $l_{\rm IR}$ is introduced.
 Therefore, the probability is peaked at $\lambda=\lambda_c$, where $\lambda_\mathrm{min}=0$ is realized:
\begin{align}
\lambda_c = 
\begin{cases}
\dfrac{1}{4SG^2}  \quad(\text{radiation})\\ \\
\dfrac{4}{27G^3M^2} \quad(\text{matter})
\end{cases}.
\end{align}

Furthermore, the coupling constants other than $\lambda$ are tuned in such a way that Eq.~\eqref{Eq:divergent_contribution} is maximized.
Therefore, the energy of the radiation $S$ and the matter $M$ are maximized for the universe with radiations and matters, respectively.
We call them maximum entropy principle~\cite{Kawai:2011rj,Kawai:2011qb,Hamada:2014ofa,Hamada:2014xra,Hamada:2015dja} and maximum matter principle.  
As a result, the fine-tuned cosmological constant, $\lambda_c$, becomes almost zero.\footnote{For the universe with radiations, by assuming that $S/l^{4/3}$ equals to the energy density of the cosmic microwave background, the fine-tuned cosmological constant, $\lambda_c$, is much smaller than the observed value. The explanation of the small but finite cosmological constant is beyond the scope of this paper.}

Let us discuss the two phenomenological implications of the maximum entropy principle. 
One is the flatness of inflaton potential.   
Assuming the instant reheating, we observe
\aln{
&S=\rho_{\rm inf}^{}a_{\rm end}^4=\rho_{\rm inf}^{}e^{4N}a_{\rm ini}^{4}~,
\label{initial entropy}
}  
where $\rho_{\rm inf}^{}$ is the vacuum energy of the inflation, $N$ is the total e-folding number, and $a_{\rm end}^{}$ ($a_{\rm ini}^{}$) is the radius of the universe at the end (onset) of inflation.   
For a given value of $\rho_{\rm inf}^{}$, $S$ is an increasing function of $N$, which means that a flatter inflaton potential is preferred.\footnote{
Of course, we need to explain why the model with finite $N$ is realized in our universe. This may require a new idea. 
}
It is noteworthy that the Higgs potential can actually have a saddle point around the Planck scale~\cite{Hamada:2013mya,Hamada:2014wna,Hamada:2021jls}, by tuning the top-quark mass.
This could be viewed as one of the signatures of the maximum entropy principle.    
 
The other implication is the strength of a strong first-order phase transition. 
When the universe undergoes a first-order phase transition, the radiation entropy increases due to the release of latent energy.  
Suppose that a first-order phase transition happens at the time $t=t_*^{}$ (and temperature $T=T_*^{}$) and that all the latent energy $\Delta V$ is converted to radiation energy. 
Then, the entropy production is
\aln{\delta S=\Delta V a^4(t_*^{})=\frac{\Delta V}{\rho_{\rm rad}^{}(T_*^{})}\rho_{\rm rad}^{}(T_*^{})a^4(t_*^{})=\alpha S_{\rm ini}^{}~,
}
where $\rho_{\rm rad}^{}(T_*^{})$ is the radiation energy density right before $T=T_*^{}$, 
$\alpha=\Delta V/\rho_{\rm rad}^{}(T_*^{})$ is the strength parameter of first-order phase transition, and $S_{\rm ini}^{}$ is the entropy before the phase transition. 
One can see that $\delta S$ linearly depends on  $\alpha$, which means that a strong first-order phase transition is preferable by the maximum entropy principle. 
The maximum matter principle may also have a lot of implications for particle physics and cosmology such as various dark matter scenarios, baryogenesis, primordial black holes, and so forth.


\section{Conclusion}
In this paper, we have studied the validity of Coleman's mechanism for fine-tuning problems in two-dimensional and four-dimensional quantum gravity theories. 
In two-dimensional Euclidean gravity, we have shown that the mechanism does not work because the effect of baby universes is too small. 
Matrix models can give alternative approaches to realize the mechanism, but their naive non-local modification also does not work since the partition function is divergent. 

As a concrete example for the realization of Coleman's mechanism, we have proposed a Lorentzian non-hermitian model of the quantum universe.  
Such a non-Hermitian property was motivated by the physical intuition that the annihilation of an universe to nothing should be highly suppressed because of the matter fields. 
We have shown that the static distribution of coupling constants Eq.~(\ref{equilibrium distribution 2}) has a very strong and non-trivial peak depending on the matter contents of the universe.  
In the case of the spatially flat universe with a   cosmological constant $\lambda$, the wave function has a peak at the point where 
$\lambda$ vanishes, and this resembles the original Coleman's baby universe theory.  
In a more realistic universe, we have shown that the distribution has a strong peak at which the entropy or matter energy becomes maximum, and we call it the maximum entropy principle or the maximum matter principle. 

There are still many open questions that should be addressed: 
In this paper, we have omitted the kinetic Hamiltonian of baby universe i.e. $\hbar \omega \hat{a}^\dagger \hat{a}$ for simplicity, but its existence can change the whole dynamics significantly.  
Moreover, the assumption of non-Hermiticity of the model was also not fully justified and we need more logical reasoning to explain that. 
We would like to study these issues in future investigations. 
%

\section*{Acknowledgements} 
We would like to thank Pablo Soler for the useful discussions.   
The work of YH was supported by JSPS Overseas Research Fellowships. 
At the final stages of the work, YH was supported by MEXT Leading Initiative for Excellent Young Researchers Grant Number JPMXS0320210099. 
HK thanks Prof. Shin-Nan Yang and his family for their
kind support through the Chin-Yu chair professorship. HK is partially supported by JSPS (Grants-in-Aid for Scientific Research
Grants No. 20K03970 and 18H03708), by the Ministry of Science and Technology, R.O.C.
(MOST 111-2811-M-002-016), and by National Taiwan University.
K.K. would like to thank Yukawa Institute for Theoretical Physics, Kyoto University for the support and the hospitality during his stay by the long term visiting program. 

\appendix

\section{Random Walk}\label{Sec:Random_Walk}
Consider a random walk on a lattice space of dimension $D$.
The number of paths consisting of $L$ steps is given by
\begin{align}
N(L) \sim (2D)^L~.
\end{align}
The number of closed paths consisting of $L$ steps is obtained by multiplying the probability of returning to the original point. From the central limit theorem, the probability density of observing the particle at position $x$ is
\begin{align}
P(x) = \frac{1}{(2\pi \sigma)^{D/2}} e^{-\frac{x^2}{2\sigma}}
\end{align}
and the variation $\sigma$ is
\begin{align}
\sigma=\kappa L,
\end{align}
where $\kappa=1/D$ for the $D$-dimensional lattice space.
Now the number of the closed paths is
\begin{align}
N_{\text{closed}}(L)\sim (2D)^L P(0)
= (2D)^L (2\pi \kappa)^{-D/2} L^{-D/2}.
\end{align}
The power dependence $L^{-D/2}$ is universal and independent of the detailed definition of a random walk.

\bibliographystyle{TitleAndArxiv}
\bibliography{Bibliography}

\end{document}